IS THE EMERGENCE OF LIFE AN EXPECTED PHASE TRANSITION IN THE EVOLVING UNIVERSE?

STUART KAUFFMAN AND ANDREA ROLI

## Abstract

We propose a novel definition of life in terms of which its emergence in the universe is expected, and its ever-creative open-ended evolution is entailed by no law. Living organisms are Kantian Wholes that achieve Catalytic Closure, Constraint Closure, and Spatial Closure. We here unite for the first time two established mathematical theories, namely Collectively Autocatalytic Sets and the Theory of the Adjacent Possible. The former establishes that a first-order phase transition to molecular reproduction is expected in the chemical evolution of the universe where the diversity and complexity of molecules increases; the latter posits that, under loose hypotheses, if the system starts with a small number of beginning molecules, each of which can combine with copies of itself or other molecules to make new molecules, over time the number of kinds of molecules increases slowly but then explodes upward hyperbolically. Together these theories imply that life is expected as a phase transition in the evolving universe. The familiar distinction between software and hardware loses its meaning in living cells. We propose new ways to study the phylogeny of metabolisms, new astronomical ways to search for life on exoplanets, new experiments to seek the emergence of the most rudimentary life, and the hint of a coherent testable pathway to prokaryotes with template replication and coding.

## Introduction

Erwin Schrödinger, in his famous 1944 book, *What is Life?*, brilliantly proposed the aperiodic crystal as the source of order in organisms,(1). But he left the question open. Eighty years later, building upon previous work, we believe a coherent picture can be drawn. The universe may hold a million trillion habitable planets. The emergence of life in the universe is a miracle, but, in our perspective, an expected one. Once life emerged, its evolution is radically creative and cannot be based on physics alone. Strong reductionism here shows its limits. Life is a double miracle.

## Part I.  A Definition of Life

We have no agreed-upon definition of life. We here build toward the following: Life is a non-equilibrium, self-reproducing chemical reaction system that achieves: i. Collective autocatalysis, ii. Constraint Closure, iii. Spatial Closure; iv. as such, living entities are Kantian Wholes. We explain these concepts below.

**Collectively Autocatalytic Sets**

A collectively autocatalytic set, CAS, is an open chemical reaction system fed with exogenous molecular and energetic building blocks, having the property that a last chemical reaction step forming each molecule in the set is catalyzed by at least one molecule in the set or by one molecule in the food set. Figure 1a shows a simple example, (2). Figure 1b shows a more complex example (3).

The concept that life must be based on template replicating polynucleotides has dominated the origin of life field for some 50 years, (4,5). Yet replication of a "nude replicating RNA gene" has not yet been achieved, (6). Nevertheless, this goal may be achieved.

The familiar concept of a template replication double stranded RNA sequence is a specific example of a collectively autocatalytic set. Each strand is a template catalyst for the synthesis of the other strand. However, the concept of collective autocatalysis is far broader.

In sharp contrast to the hopes for a replicating RNA sequence, collectively autocatalytic sets of DNA, of RNA, and of peptides have been constructed. The first, a DNA collectively autocatalytic set, was constructed by G. von Kiedrowski, (7). An RNA collectively autocatalytic set was achieved by N. Lehman and colleagues, (8). This set spontaneously self organizes given its building blocks. A collectively autocatalytic set of nine peptides constructed by G. Ashkenasy, (9), is shown in Figure 2. Autocatalytic sets of lipids have also been considered, (10).

These results are of fundamental importance. Creating self-reproducing open chemical reaction systems is achieved.

Stunning evidence now demonstrates the presence of small molecule collectively autocatalytic sets containing no DNA, RNA, or peptide polymers, in all 6700 prokaryotes, Figure 3, (11,12). These small molecule self-reproducing sets contain from hundreds to several thousand small molecules and reactions among them. These autocatalytic sets synthesize several amino acids and ATP. The sets are identified computationally. It remains to be shown that they reproduce in vitro.

The presence of small molecule collectively autocatalytic sets in all 6700 prokaryotes strongly suggests that the first chemical systems capable of self-reproduction in the universe were precisely such sets. We show below that the emergence of such sets is expected.

The identification of small molecule autocatalytic sets bears on the ongoing debate concerning the necessity for template replicating polynucleotides in the origin of life. Such a "nude RNA gene" would have to evolve RNA sequences to catalyze a connected metabolism to create and sustain its own building blocks. However, there is no reason at all why such a connected metabolism by itself would also be collectively autocatalytic. This consideration increases the confidence that the origin of molecular reproduction was through the emergence of small molecule collectively autocatalytic sets.

**Life: Kantian Wholes, Catalytic Closure, Constraint Closure, Spatial Closure**

*Kantian Wholes*

In the 1790s, philosopher Immanuel Kant introduced a fundamental concept: An organized being has the property that the *Parts* exist for and by means of the *Whole*, (13). Kant's insight has lain dormant for 230 years. All living beings are Kantian Wholes that exist for and by means of their Parts. You are a *Kantian Whole*. You exist by means of your Parts – heart, liver, kidneys, lungs, brain. Your Parts exist by means of you, the Whole. You reproduce, and your children inherit your Parts.

All living organisms are Kantian Wholes. This includes the doubted class of viruses. Inside the environment of the cell, viruses are Kantian Wholes that reproduce. The Parts of the virus, in the context of the cell, create multiple copies of the Parts of the virus that self-assemble into the mature virus Whole. It is of interest that a definition of life including that of a Kantian Whole classifies viruses as alive.

Kantian Wholes are a special class of dynamical physical systems. A crystal is not a Kantian whole. The atoms of the crystal can exist without being parts of the crystal. A brick is not a Kantian Whole. A cell is a Kantian Whole.

*Catalytic Closure*

A collectively autocatalytic set, such as the 9-peptide set in Figure 2, achieves *Catalytic Closure*. Each reaction in the system is catalyzed by at least one molecule in the system. All living cells achieve catalytic closure. No molecule in a living cell catalyzes its own formation. The set molecules in a living cell, a Whole, achieves catalytic closure as the cell reproduces, (14,15,16).

Systems that achieve catalytic closure are also Kantian Wholes. Each of the peptides in the 9-peptide collectively autocatalytic set in Figure 2 is a Part that exists for and by means of the Whole set of nine peptides whose mutual catalysis enables all the Parts to exist.

*Constraint Closure*

Living cells, including a small molecule collectively autocatalytic set of the type found in all 6700 prokaryotes, achieve a newly recognized and profound property: *Constraint Closure,* (17). Thermodynamic *work* is the constrained release of energy into a few degrees of freedom, (18). An example is a cannon with powder at its base and a cannon ball nestled next to the powder. When the power explodes, the cannon, that is both a boundary condition and a constraint, constrains the release of energy to blast the cannon ball down the bore of the cannon. Thermodynamic work is done on the cannon ball. Therefore, in the absence of constraints on the release of energy in a non-equilibrium process, no thermodynamic work can be done, (19).

Newton does not tell us where the boundary conditions come from. The cannon in the example is the boundary condition. But where did the cannon come from? The critical answer is that thermodynamic work was required to *assemble* the cannon. We may conclude: No Constants, No Work. But it often takes work to construct the relevant constraint. Hence: No Constraints, No Work. No Work, No Constraints. This Work-Constraint cycle is a new issue, (19).

Maël Montévil and Matteo Mossio in 2015 first defined Constraint Closure, (17):

Consider a system with three non-equilibrium processes, 1, 2, and 3. Consider three constraints, A, B, and C. Let A constrain the release of energy in process 1 to construct a B. Let B constrain the release of energy in process 2 to construct a C. Let C constrain the release of energy in process 3 to construct an A (see Figures 1a, 1b and 2).

The above system achieves a remarkable property: *Constraint Closure.* The set of constraints, here A, B, and C, constrain the release of energy of a set of processes, here 1, 2 and 3, into the few degrees of freedom that therefore do thermodynamic work construct the very same set of constraints, A, B, and C! This system literally does the thermodynamic work to construct itself by constructing its own boundary condition constraints on the release of energy that construct the same boundary conditions.

Constraint closure is an entirely new concept. We construct our automobiles. An automobile is an elaborate arrangement of parts that constrain the release of energy of parts that impinge on other parts. Gas explodes, pistons move, wheels turn. But automobiles do not construct their own boundary condition constraints on the release of energy.

All collectively autocatalytic molecular reaction systems achieve both catalytic closure and constraint closure. All are Kantian Wholes. For example, in the 9-peptide collectively autocatalytic set in figure 2, each peptide acts as a ligase by binding the two fragments of the next peptide. By orienting the two fragments, the peptide as a ligase lowers the activation barrier to the ligation of the two fragments to make a second copy of the next peptide. Thermodynamic work is done to construct the next peptide as a peptide bond is formed. Because this is true of all the reactions in this collectively autocatalytic peptide system, the system – as a Whole – achieves both Catalytic Closure and Constraint Closure. The system constructs itself. And the system is also a Kantian Whole.

It is of the deepest importance that all living cells achieve constraint closure. Cells construct the very boundary conditions on the release of energy that constructs the very same boundary conditions. Cells construct themselves. Computers and locomotives do not construct themselves.

Reproducing cells are fundamentally *not* von Neumann's self-reproducing automata, (20). These are based on a "Universal Constructor." To construct anything specific, the Universal Constructor requires specific "Instructions." These are encoded in a physical system placed inside the Universal Constructor. The physically embodied *instructions play dual roles:* They are *used* to construct a copy of the Universal Constructor into which a *physical copy* of the physical Instructions is constructed and then inserted. The dual roles of the physical Instructions constitute precisely the distinction between software and hardware. In sharpest contrast, a living cell, via catalytic and constraint closure, constructs specifically itself. A cell is not a universal constructor requiring separate Instructions. The self-reproducing 9-peptide set in figure 2 has no separatable "Instructions" that encode its formation. The concepts of software and hardware here are void.

Paul Davies, (21), points out that, in the context of a living cell, genes together with the transcription and translation apparatus are, in fact, a universal constructor for all possible encoded polypeptides. The genes can be regarded as a set of instructions. However, the living cell in which the genes are located is not itself a universal constructor. It constructs specifically itself. Were each of its several thousand genes substituted by a random DNA sequence encoding some random polypeptide, the cell synthesizing these novel proteins would almost surely perish.

Living organisms have evolved to form nested Kantian Wholes. A prokaryote is a first order Kantian Whole. A eukaryotic cell, a symbiont with mitochondria and chloroplasts, (22, 23), is a second order Kantian Whole containing first order Kantian Wholes. A multi-celled organism is a third order Kantian Whole containing second order and first order Kantian Wholes.

## Part II. The first Miracle: The emergence of life is an expected phase transition – TAP and RAF.

In this Part III we show that the emergence of collectively autocatalytic sets is an expected phase transition in chemical reaction networks as the diversity of molecular species in the system increases, and the diversity of the reactions among them increases even faster. We discuss first the phase transition to collective autocatalysis, RAF, in sufficiently rich chemical reaction networks. Then we marry the RAF theory to the TAP theory that yields the increasing diversity of chemical species in which the RAF phase transition must eventually occur. This TAP RAF union is new.

The molecules of life are combinatorial objects made of a diversity of atoms, CHNOPS, Carbon, Hydrogen, Nitrogen, Oxygen, Phosphorus, Sulfur, (24). A molecule comprised of, say 10 atoms, each bonded to one or two other atoms, can be constructed from a diverse set of fragments of that molecule. This diverse set is the set of substrates to reactions by which it can be formed from its components.

A simple example is a linear polymer of two building blocks A and B. An example is (ABBABAABBA). This polymer has ten building blocks. This single polymer can be constructed in 9 different ways, as is easily seen by breaking any one of the 9 bonds between adjacent building blocks. As the combinatorial complexity of molecules in a system increases, the ratio of reactions, R, to molecules, M, R/M, in the total reaction system *increases*.

A chemical reaction graph is a "bipartite graph" consisting of dots representing the species of molecules, and boxes representing the different reactions by which molecules transform into each other. Arrows point from the dots representing the substrates of a reaction into the reaction box. Arrows point from the reaction box to the products of the reaction. This graph represents the structure of the chemical reaction network, not the thermodynamic direction of flow that depends upon the displacement from chemical equilibrium, (3, 15).

There is a critical next step: If we knew which molecules catalyzed which reaction, we could represent this by a dotted arrow from the relevant catalyst molecule to the reaction it catalyzed. This structure is called a bipartite hypergraph, see figure 1b. Given a specific bipartite hypergraph it can be examined to see if it contains a collectively autocatalytic set. The reaction system in Figure 1b is an example of such a bipartite hypergraph that contains a collectively autocatalytic set, (3).

In general, we do not know which molecule catalyzes which reaction. In the absence of specific knowledge, theory and insight can proceed by the naïve assumption that each molecule has some probability, Pcat, to catalyze each reaction. This simple assumption already leads to remarkable results. Collectively autocatalytic sets arise as a first order phase transition as the diversity of molecules and reactions among them increase. We see why next, (3, 15,25).

**Random Graph Theory**

In 1959, two mathematicians, Erdos and Renyi, published a seminal paper on the properties of random graphs, (26). A graph is a set of dots or nodes or vertices. Each dot may be connected by a line to no other dots, one other dot, or some number of other dots.

Erdos and Renyi asked a wonderful question: Start with N nodes. Randomly pick up a pair of nodes and connect them by a line. Iteratively keep picking up random pairs of nodes and connecting them with lines.

Let N be the number of Nodes. At any step in this process let the number of lines connecting nodes be L. Consider the ratio: L/N. What happens to the graph as L increases for fixed N?

Magic happens when the ratio of L/N increases to 0.5. Suddenly a giant connected component, or web, arises. In such a giant component, each node is directly or indirectly connected to all the other nodes in that giant component. The ratio L/N = 0.5 is a *first order phase transition.* Given a fixed number of nodes, N, as L increases and passes L/N = 0.5, almost certainly a giant connected component arises.

**The Emergence of Molecular Reproduction as a First Order Phase Transition**

The emergence of collectively autocatalytic sets, also called RAFs, arises as the same phase transition in bipartite chemical reaction hypergraphs. Consider a given bipartite chemical reaction graph. Consider increasing the probability, Pcat, that any molecule catalyzes any given reaction. For each value Pcat assign at random according to Pcat which molecules catalyze which reactions. Does the system contain a collectively autocatalytic set? At some value of Pcat so many reactions are catalyzed that they form a giant component that is now collectively autocatalytic, (3,15,26).

This is precisely the first-order phase transition to molecular reproduction in sufficiently rich non-equilibrium chemical reaction systems.

*More importantly, keep Pcat constant and increase the number and atomic complexity of the molecules in the system.* The ratio of reactions to molecules, R/M, must increase. At some complexity of the molecules in the system, and the ratio R/M, a collectively autocatalytic set, an RAF, will emerge with probability approaching 1.0, (3). This is the essential *first-order phase transition* by which self-reproducing molecular systems such as the small molecule collectively autocatalytic sets in all 6700 prokaryotes can have arisen.

**The History of an Idea**

One of us, (14), created the first model demonstrating the spontaneous emergence of collectively autocatalytic sets in 1971. Two publications emerged in 1986, one by Kauffman alone with theorems demonstrating that such an emergence was expected, (15). The second by Farmer, Packard and Kauffman, (3), implemented a code based on a "binary polymer model". Each model polymer was comprised of two monomers, A and B. The maximum length of a polymer, L, was fixed. All $2^{(L+1)}$ possible polymers could be part of the system. The reaction system was sustained by a "food set" consisting in a subset of the possible polymers. Only cleavage and condensation reactions were allowed. Each polymer had the same fixed probability, Pcat, to catalyze each reaction.

The results were convincing: For a fixed Pcat, as the length of the longest polymer allowed, L, increased, so the diversity of potential polymers increases as did the R/M ratio, collectively autocatalytic sets arose with probability approaching 1.0. Figure 1b is one such collectively autocatalytic set.

Extensive work over the next decades, (27,28, 29), has demonstrated: i. The rules assigning at random which polymer catalyzes which reaction can be uniform or power law. This makes little difference. ii. Models in which sequence recognition is included make little difference. iii. The number of reactions each polymer must catalyze is between 1 and 2. This is chemically plausible. iv. Importantly, Hordijk and Steel found that such collectively autocatalytic sets contain "irreducible autocatalytic sets" that jointly form more complex systems of many such irreducible sets, (28, 29). v. Vassas et al. pointed out that such irreducible autocatalytic sets, each able to function as an independent replicator, could function as "genes" that could be inherited and partitioned to daughter sets. Thus, such simple systems with no "genome" can evolve, (29). Such an exchange inheritance of irreducible autocatalytic sets almost surely played a large role in the early evolution and phylogeny of the small molecule autocatalytic sets now found in all 6700 prokaryotes, (11,12). Irreducible autocatalytic sets can be identified computationally. Therefore, such an evolution of metabolism is now open to detailed study.

**The Expected Emergence of Molecular Reproduction in the Evolving Universe**

A first-order phase transition to molecular reproduction is expected in the chemical evolution of the universe where the diversity and complexity of molecules increased. At the earliest stage there were the fundamental particles, quarks, gluons, electrons, positrons. As the universe cooled, hadrons formed. Then the first elements, hydrogen, beryllium, formed. Later, in supernovae, the rest of the 98 stable atoms formed, (30).

The emergence of simple then ever-more complex molecules followed the same pattern from simple molecules and low diversity upward. The diversity, atomic complexity of the molecules, and the potential reactions among them increased, (31). The Murchison meteorite, formed five billion years ago with our solar system, has hundreds of thousands of molecular species and potential reactions among them, (32).

The theory we discuss here predicts that at some sufficient diversity of molecular species, M, and reactions among them, R, hence a sufficiently large R/M ratio, collectively autocatalytic sets will emerge as a first order phase transition.

**The New Mathematical Theory – TAP and RAF**

The theory of the emergence of collectively autocatalytic sets, RAFs, is well established, (15,16,25,27,28,29). We here marry that RAF theory to an independent theory, TAP, (33,34,35), that can explain the increasing diversity of molecular species in the evolving universe.

*The TAP Equation*

The TAP equation and its behavior are shown in Figure 4.

The dynamical properties of this simple equation are now well studied numerically and with theorems, (33,34,35). The properties are remarkable. The system is a discrete dynamical system in which molecules can combine with each other to form new molecules. If the current number of molecules is $M_t$, then for each subset of size i chosen among the $M_t$, ($M_t$ choose i), the equation combines them to create a new molecule with a probability $alpha \wedge i$, for $0 < alpha < 1.0$.

If the system starts with a small number of beginning molecules, each of which can combine with copies of itself or other molecules to make new molecules, over time the number of kinds of molecules increases slowly then explodes upward hyperbolically and reaches infinity in a finite time, (33,34 35), Figure 4.

The TAP process is a crude model of the increasing chemical diversity of the universe, (31,32).

We now unite TAP and RAF. This union hopes to explain the expected emergence of collectively autocatalytic systems as a first order phase transition in the evolving universe. The simple step is to allow each molecule in TAP to catalyze each reaction in TAP at random with a fixed probability Pcat. The first results with respect to technological evolution were just published, (36). As time passes, the diversity of entities increases, then the first order phase transition to the emergence of collectively autocatalytic sets arises with probability 1.0, (36).

This united TAP-RAF theory demonstrates a basic truth. In the chemical evolution of the universe, molecular diversity increased by some analogue of the TAP process. As this occurred, the complexity of molecules increased, thus the number of reactions increased as did the ratio of reactions, R to molecules, M, R/M. But the same set molecules, M, are candidates to catalyze the same set of reactions, R, among the set of M molecules. Therefore, with any rough probability of catalysis, assigned among the molecules uniformly, as a power law, or otherwise, at some point the first order phase transition arises. Molecular reproduction is expected to arise in the evolving universe.

**The Spontaneous Emergence of Élan Vital – Evolving Life**

Any such non-equilibrium reproducing molecular reaction system is a Kantian Whole that achieves Catalytic Closure. In addition, the molecules that serve as catalysts are, themselves, boundary condition constraints of the release of energy in the specific reactions they catalyze. Thus, the system constructs its own boundary condition constraints on the release of energy that constructs the same boundary condition constraints. The system achieves Constraint Closure. We add some form of enclosure, for example, in a tiny pocket in a hydrothermal vent, or better, in a liposome whose lipids are synthesized by the same system, (37,38).

This newly recognized union of four closures, Kantian Whole, Catalytic Closure, Constraint Closure, and Spatial Closure is Bergson's mysterious élan vital, (39), here rendered entirely non-mysterious. These four conditions constitute life. *Life arises as an expected phase transition in the evolving universe.*

We note parallels to recent work by R. Hazen and colleagues suggesting a new law for the emergence of functional diversity in the abiotic world, (40). The authors consider a set of stable objects, such as a diversity of molecules, and stable transformations among them, such as chemical reactions among the molecules, and point out that complex systems such as a high diversity of abiotic minerals form. "Function" is defined for features of minerals. These authors do not consider the emergence of Kantian Wholes that are collectively autocatalytic sets achieving catalytic closure and constraint closure, the emergence of evolving life. Here, "function" is defined in terms of how Parts evolve to sustain the evolving Whole.

## Part III. The Second Miracle: The evolution of the biosphere is a propagating, non-deducible construction, not an entailed deduction. There is no Law. Evolution is ever-creative.

All classical physics, the physics of Newton, lives within the Newtonian Paradigm: i. State the relevant variables, for example position and momentum. ii. State the Laws of Motion in differential form connecting the relevant variables. Newton's three laws of motion are an example. iii. Define the boundary conditions. These define the phase space of all possible combinations of values of the relevant variable. iv. State the initial conditions. v. Integrate the equations of motion to obtain the entailed determined single trajectory in the system's phase space, (41,42,43).

The Newtonian Paradigm is unchanged in Quantum Mechanics. Schrödinger's wave equation is integrated to obtain the entailed trajectory of a probability distribution. Then measurement, typically held ontologically random, occurs, (42,43).

It is fundamental to the entire Newtonian paradigm, hence of all modern physics, that *the phase space must be specified beforehand,* (42,43).

We show next that the evolving biosphere of Kantian Wholes persistently creates novel phase spaces that cannot be deduced or determined ahead of time. The entire Newtonian paradigm collapses. The evolving biosphere cannot be explained by physics alone. Appeal to "function" is necessary.

These issues are basic, (42,43):

i. Once we have defined a Kantian Whole, the non-circular definition of the "function" of a Part is clear. The function of a part is *that subset* of its causal consequences that sustains the Whole. The function of your heart is to pump blood, not make heart sounds or jiggle water in your pericardial sac.

ii. Selection acts at the level of the Kantian Whole organism, not its Parts. Selection does not directly select for hearts that are more efficient at pumping blood. Organisms that inherit such improved hearts are more likely to have offspring, thus improved hearts are indirectly selected.

iii. Because the function of a Part is that *subset* of its causal properties that sustains the Whole, the *function of the very same part can change*. Some new, unused, subset of causal properties of the same Part can come to sustain the Whole. These are called Darwinian pre-adaptations, or by Gould and Verba, exaptations, (44). Examples include the co-opting of scales that evolved for thermoregulation on some dinosaurs to evolve into flight feathers on birds. Other examples include the co-opting of normal enzymes to become transparent lens proteins. A superb example is the evolution of swim bladders from the lungs of lung fish, (43).

iv. A remarkable and fundamental feature of such exaptations is that they cannot be deduced. Consider a hypothetical example. An engine block can be used as a paper weight. The same engine block has sharp corners that can be used to crack open a coconut. But from the fact that an engine block is being used as a paper weight, it cannot be deduced that "This object can be used to crack open coconuts". This object might have been a banana peel.

v. Such new uses of the same object are "Jury Rigging." There is no deductive theory of Jury Rigging.

vi. The truly profound implication is that such *non-deducible jury-rigged exaptations are the source of functional novelty and the open-ended evolution of the biosphere*. The evolution of the biosphere is a non-deducible construction not an entailed deduction. No entailing laws govern the evolution of the biosphere, (42,43).

vii. The further implication is equally profound. We cannot list all the uses of an engine block alone or with other things. We also cannot list all the uses of a screwdriver alone or with other things. *Therefore, we cannot use Set Theory or any mathematics based on set theory:* The First Axiom of Set Theory is the Axiom of Extentionality, (45): "Two sets are equal if and only if they contain the same

- viii. *members". But we cannot prove that the un-listable uses of an engine block are identical to the un-listable uses of a screwdriver. More the Axiom of Choice fails as well. The implication is huge: We can use no mathematics based on Set Theory – essentially all of mathematics – to deduce the future evolution of the biosphere.*
- viii. *No union and intersection of Sets. No First Order Logic. No numbers via Russell, (43, 46). No numbers via Peano, (43, 47). No equations. No real numbers. No real line. No manifolds. No topology. No equations on manifolds. No integration of the equations we cannot write.*
- ix. The entire Newtonian Paradigm that is the basis of all physics, requires a prestated phase space. But we can neither deduce nor prestate the evolving phase space of the evolving biosphere, (42,43). Evolution is beyond the Newtonian Paradigm.
- x. We cannot explain the evolving biosphere with physics alone. The heart evolved by virtue of its function pumping blood. Natural selection of heritable variation acts on the Kantian Whole, not directly on its Parts. Such selection is Downward Causation. Here, the explanatory arrows point upward.
- xi. Strong reductionism, the dream of many, fails. If the Final Theory to be inscribed on the famous T-shirt is to include the deduction of the evolving biosphere, there is no final theory.
- xii. Heisenberg's demonstration of the Uncertainty Relation demanded that we abandon Classical Physics, (48).
- xiii. If we must abandon Set Theory with respect to biological evolution, what are the implications? We hardly begin to know.
- xiv. We now enter the Third Transition in Science, (43). The new fundamental questions surely include: How does the evolving biosphere create, and seize by heritable variation and Natural Selection, or genetic drift, the ever-new bubbles of possible ways organisms can co-exist for some period as the burgeoning wave of life flowers onward.

## Part IV.  New Observations and Experiments: Is There Life in the Cosmos?

- i. We are discovering ever more exoplanets, (49,50). We seek evidence of life on these planets, (50). In addition, we seek evidence of life on Mars and elsewhere in the Solar System, (51,52,53). The results discussed here suggest a potent new hope. Small molecule autocatalytic sets are likely to be the simplest forms of life in the universe. They are Kantian Wholes achieving catalytic and constraint closure. If true, a search for life in the cosmos can include search for small molecule collectively autocatalytic sets.

    If we establish that the small molecule collectively autocatalytic sets in all 6700 prokaryotes do reproduce chemically in vitro, we can then seek evidence in the Solar System and beyond for just such small molecule collectively autocatalytic sets. The total number of reactions observed among the 6700 prokaryotes is 5994, (11,12). The sizes of these sets range from 3 to 600 reactions. In general, the large sets each contain smaller "irreducible collectively autocatalytic sets", (11,12). We can seek for these in the chemistry of ancient mudstones on Mars**, (**51), on Enseladus, (52), and elsewhere, (53). More specifically, we may find evidence of

such small molecule autocatalytic sets. Even more, we may find evidence that the molecules in the catalytic core of such sets are present at above equilibrium concentrations with respect to other molecular inputs to the set. This latter discovery would strongly suggest that such systems were functioning as non-equilibrium autocatalytic networks: Life on Mars.

ii. It is now feasible to create high diversity small molecule libraries. For example, work by Ott running the Miller Urey experiment starting with only four molecular species for a month yields thousands of small molecules, identified by mass spectrometry, (54). We can now ask if in such systems, small molecule collectively autocatalytic sets can emerge. This is a "Go or No Go" experiment. If No, the theory is probably wrong.

iii. If Yes, we can begin to envision testable pathways beyond small molecule collectively autocatalytic sets to such sets becoming the metabolism of peptide RNA autocatalytic sets with which they co-evolve as new Kantian Wholes. In turn these might evolve to template replication, and even to genetic coding, (55,56,57). Real experiments are needed.

iv. We begin to envision testable pathways from the earliest small molecule Kantian Wholes to the emergence of prokaryotes. From these to the eukaryote may be a long step, but multicellularity arose six or more times. Is the emergence of complex life so very improbable? We do not know. Perhaps not.

## Conclusion

We have sought the source of life in all our creation myths among all the peoples of the earth, perhaps back to Neanderthal 500,000 years ago. The issue of the Origin of Life as a scientific problem arose with Pasture's claim: Life only comes from life, (58).

Experimental efforts have been underway since Haldane and Oparin, (59,60), then the famous Miller Urey experiments in 1953, (61). Intense efforts based on the conviction that life must be based on template replication of polynucleotides have been carried out, (4,5,6). So far, no case of molecular reproduction has been found on this sensible pathway.

The concept of the emergence of collectively autocatalytic sets was first formulated in 1971, (14). Such reproducing chemical sets of DNA, of RNA, and of peptides have been engineered using evolved polymer sequences. Xavier has now found small molecule collectively autocatalytic sets containing no DNA, RNA or peptide polymers in all 6700 prokaryotes, (11,12). This strongly suggests, but does not prove, that such systems may be the earliest form of molecular reproduction in the universe.

Well established mathematical theory demonstrates that such systems can arise as a first-order phase transition in a universe where the diversity and atomic complexity of molecules increases. If this is correct, as we claim, the emergence of life is an expected phase transition in the evolving universe.

Since Newton, basic science has rested on the powerful Newtonian Paradigm. This paradigm requires a prestated and knowable phase space of all the values of the relevant variables. But living organisms are Kantian Wholes that achieve Catalytic Closure, Constraint Closure, and Spatial Closure. Stunningly, we can use no mathematics based on Set Theory – all of mathematics it seems – to deduce the ever-creative evolution of the biosphere.

Life is an expected miracle in the universe whose ways of becoming are literally numberless. The 20$^{th}$ Century saw the emergence of the atomic Age, the mushroom cloud, and mutually assured destruction. With Gödel, the 20$^{th}$ century also saw the End of Certainty, (62). In this, the first quarter of the 21$^{st}$ Century, we begin barely to glimpse the astonishing blossoming creativity of the biosphere of which we are members. We are, truly, Of Nature, not Above Nature.

## Acknowledgments

We are truly grateful to Ingemar Ernberg, Carlos Gershenson, Wim Hordijk, Niles Lehman, and Ricard Solé for advice and comments.

**FIGURES**

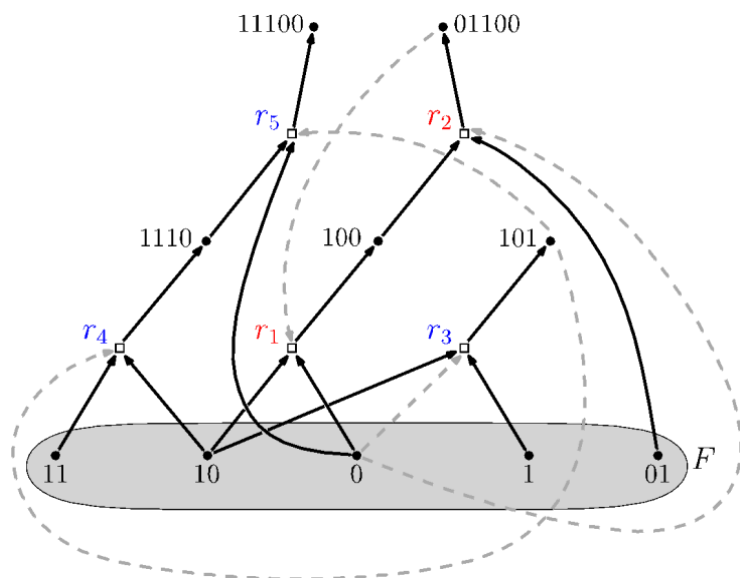

A simple collectively autocatalytic set. The model molecules are bit strings acting as substrates and products of reactions. Black solid arrows are drawn from the dots representing substrates of a reaction to a box representing the reaction. Black solid arrows are drawn from the reaction box to the dots representing the products of the reactions. The actual direction of flow of the reaction depends upon displacement from equilibrium. Dashed lines from dots representing molecules to the boxes representing reactions depict which molecules catalyze which reactions.

The exogenously supplied food set of monomers and dimers is shown in the grey oval. Derived from (28).

Figure 1a.

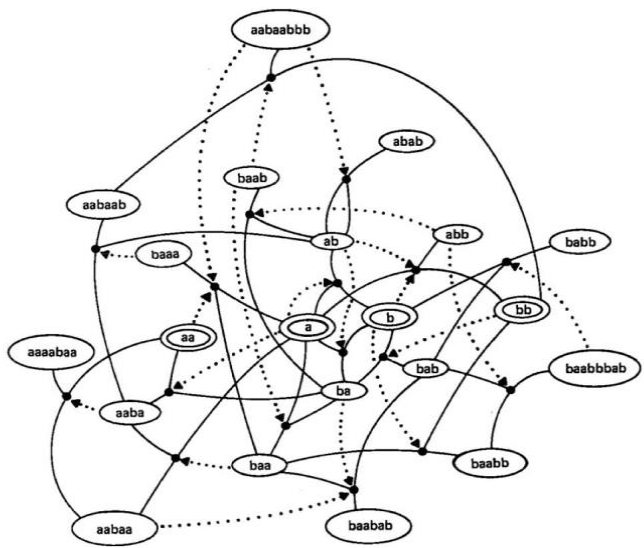

A collectively autocatalytic set of linear polymers derived from reference (3). Ovals contain polymers of two monomer types, A and B. Allowed reactions, shown as dots, are cleavage and ligation reactions. A dotted arrow from a molecule oval to a reaction dot indicates that that molecule catalyzes that reaction. Derived from (3).

Figure 1b.

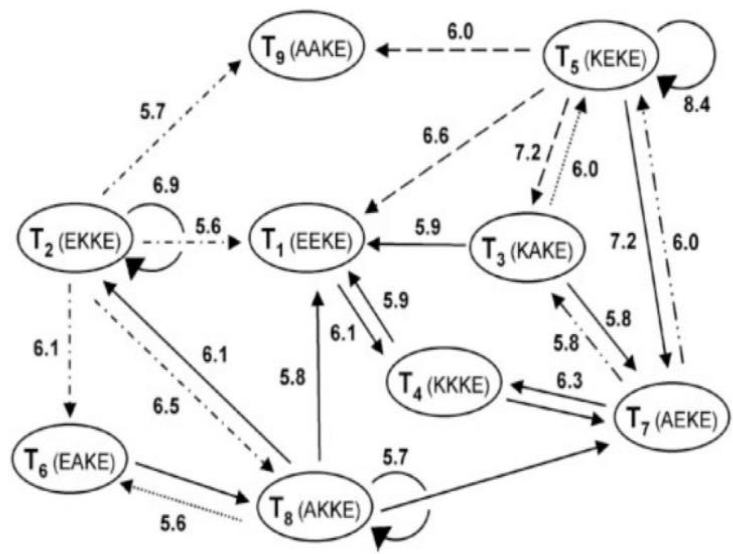

The nine peptide collectively autocatalytic set discussed in reference (9). The ovals show the molecules, the arrows show the transitions among the molecules and the relative rates.

Figure 2.

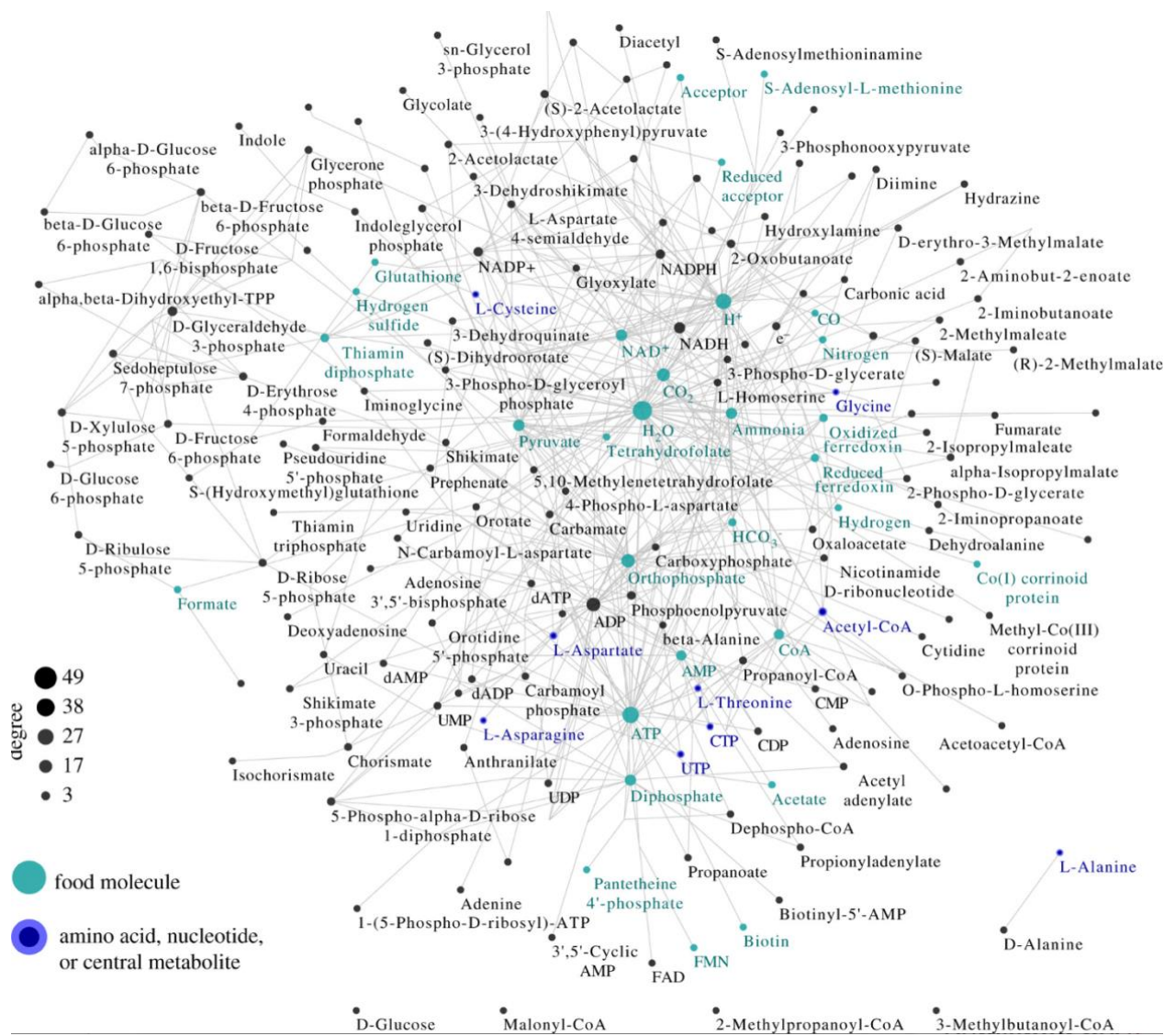

A small molecule collectively autocatalytic set with no DNA, RNA, or peptide polymers in a prokaryote. Similar small molecule autocatalytic sets are found in all 6700 prokaryotes. Presumably the phylogeny among these is part of the evolution of metabolism.

Figure 3. Derived from (11).

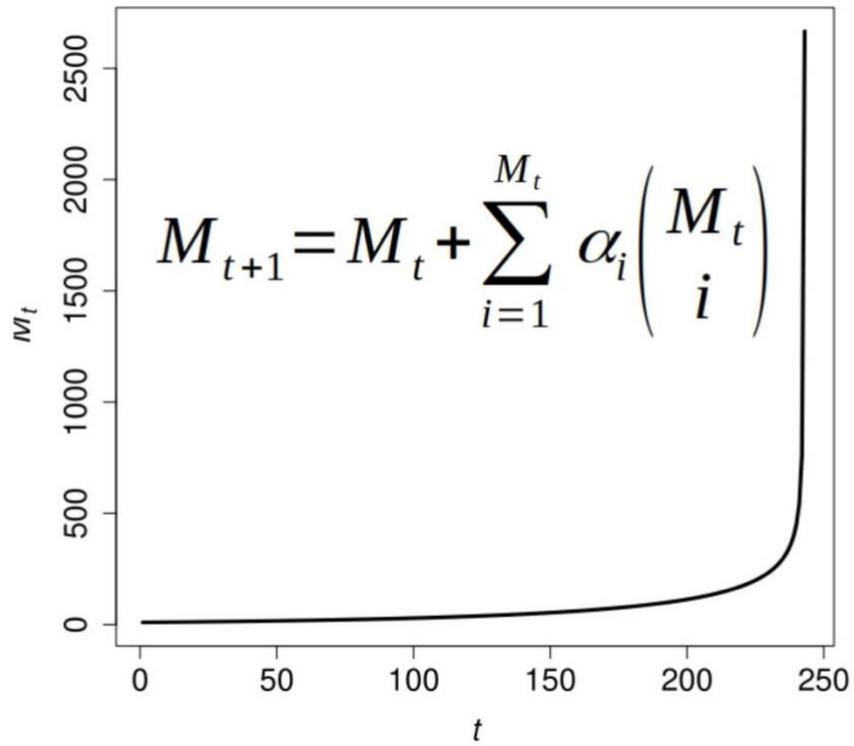

**Hockey stick growth**

The Theory of the Adjacent Possible (TAP) equation and its dynamics. Thanks to W. Hordijk.

Figure 4.

# References

1) Schrödinger, E. (1992) *What is life?: With mind and matter and autobiographical sketches*. Cambridge University Press.
2) Hordijk, W. (2013) Autocatalytic sets: From the origin of life to the economy. *BioScience* 63, no. 11, 877-881.
3) Farmer, J. D., Kauffman, S. A. and Packard, N. H. (1986) Autocatalytic Replication of Polymers. *Physica D* **2**, 50-67.
4) Orgel, L. E. The origin of life on the earth. (1994) *Scientific American* 271, no. 4 76-83.
5) Paul, N. and G. F. Joyce. (2004) Minimal self-replicating systems. *Current opinion in chemical biology* 8, no. 6, 634-639.
6) Szostak, J. W. (2012) The eightfold path to non-enzymatic RNA replication. *Journal of Systems Chemistry* 3, 1-14.
7) Sievers, D. and von Kiedrowski, G. (1998) Self-replication of hexadeoxynucleotide analogues: autocatalysis versus cross-catalysis. *Chemistry–A European Journal* 4, no. 4, 629-641.
8) Vaidya, N., Manapat, M. L., Chen, I. A., Xulvi-Brunet, R., Hayden, E. J. and Lehman, N. E. (2012) Spontaneous network formation among cooperative RNA replicators. *Nature* 491, no. 7422, 72-77.
9) Wagner, N. and Ashkenasy, G. (2015) How catalytic order drives the complexification of molecular replication networks. *Israel Journal of Chemistry* 55, no. 8, 880-890.
10) Lancet, D., Zidovetzki, R. and Markovitch, O. (2018) Systems protobiology: origin of life in lipid catalytic networks. *Journal of The Royal Society Interface* 15, no. 144, 20180159.
11) Xaiver, J., Hordijk, W., Kauffman, S. A., Steel, M., Martin, W. (2020), Autocatalytic chemical networks at the origin of metabolism *Proc Roy Soc B*, 287: 20192377. https:/doi.org/10.1098/rspb.2019.2377
12) Xavier, J. and Kauffman, S. A., (2022). Small-molecule autocatalytic networks are universal metabolic fossils. *Proc Roy Soc A*. 380:20210244
13) Kant, I. (2017) *Critique of Judgement by Immanuel Kant-Delphi Classics (Illustrated)*. Vol. 11. Delphi Classics.
14) Kauffman, S. A. (1971). Cellular Homeostasis, Epigenesis, and Replication in Randomly Aggregated Macromolecular Systems. *Journal of Cybernetics* **1**, 71- 96.
15) Kauffman, S. A**.** (1986). Autocatalytic Sets of Proteins. *J Theor Biol* **119**, 1-24.
16) Hordijk, W., Hein, J. and Mike Steel. (2010) Autocatalytic sets and the origin of life. *Entropy* 12, no. 7 1733-1742.
17) Montévil, M. and Mossio, M. (2015) Biological organisation as closure of constraints. *Journal of theoretical biology* 372 179-191.
18) Atkins, P.W. (1984) *The Second Law*, Scientific American Library, NY.
19) Kauffman, S. A. (2000) *Investigations*, Oxford University Press, NY.
20) Von Neumann, J. (1966) *Theory of self-reproducing automata*. Arthur W. Burks, ed. University of Illinois Press.
21) Davies, P. (2019) *The Demon in the machine: How hidden webs of information are solving the mystery of life*. University of Chicago Press.
22) Margulis, L. (1971) Symbiosis and evolution. *Scientific American* 225, no. 2, 48-61.

23) Margulis , L. and Sagan, D. (2000) *What is Life?* University of California Press.
24) Campbell, N. A., Williamson, B., Heyden, R. (2006) *Biology: Exploring Life*. Boston, Massachusetts: Pearson Prentice Hall.
25) Hordijk, W., and Steel, M. (2018) Autocatalytic networks at the basis of life's origin and organization. *Life* 8, no. 4, 62.
26) Erdős, P., and Rényi, A. (1960) On the evolution of random graphs. *Publ. math. inst. hung. acad. sci* 5, no. 1, 17-60.
27) Hordijk, W., and Steel, M. (2012) Autocatalytic sets extended: Dynamics, inhibition, and a generalization. *Journal of Systems Chemistry* 3, no. 1, 1-12.
28) Hordijk, W, Steel, M., Kauffman, S. A. (2012).The structure of autocatalytic sets: evolvability, enablement, and emergence. *Acta Biotheoretica*, Vol 60, Issue 4, 379-392.
29) Vassas. V., Fernando, C., Santos, M., Kauffman, S., Szathmary, E. (2012) Evolution Before Genes. Biology Direct. http://www.biology-direct.com/content/7/1/1
30) Puy, D. and Signore, M. (2002) From nuclei to atoms and molecules: the chemical history of the early universe. *New Astronomy Reviews* 46, no. 11, 709-723.
31) Kauffman, S. A., Jelenfi, D. Vattay, G. (2020) Theory of chemical evolution of molecule compositions in the universe, in the Miller-Urey experiment and the mass distribution of interstellar and intergalactic molecules *J. Theor. Biol.* Vol 486, https://doi.org/10.1016/j.jtbi.2019.110097.
32) Schmitt-Kopplin, P., Gabelica, Z., Gougeon, R., Hertkorn, N. (2010) High molecular diversity of extraterrestrial organic matter in Murchison meteorite revealed 40 years after its fall. *Proceedings of the National Academy of Sciences* 107, no. 7, 2763-276.
33) Steel,M., Hordijk W., Kauffman, S. A. (2020), Dynamics of a birth--death process based on combinatorial innovation, *J. Theor. Biol* 491, 11018.
34) Cortês, M., Kauffman, S. A., Liddle, A. R., Smolin L, Biocosmology and the theory of the adjacent possible. (2022) https://www.biocosmology.earth arXiv https://arxiv.org/abs/2204.14115
35) Koppl, R., Gatti, R., Deveraux, A., Fath, B., Herriot J., Hordijk, W., Kauffman S. A., Ulanowitcz, R., Valverde, S. (2023) *Explaining Technology*, Cambridge University Press, UK.
36) Hordijk, W. Koppl, R., Kauffman, S. A. (2023) Emergence of Autocatalytic Sets in a Simple Model of Technological Evolution, Journal of Evolutionary Economics, https://doi.org/10.1007/s00191-023-00838-2
37) Lane, N. (2015) *Vital Question: Energy, Evolution, and the Origins of Complex Life*. WW Norton & Company.
38) Damer, B., and Deamer, D. (2020) The hot spring hypothesis for an origin of life. *Astrobiology* 20, no. 4, 429-452.
39) Bergson, H. (1907). *Creative Evolution*. East India Publishing Company, Ottawa, Canada, 2022.
40) Wong, M., Cleland, C., Arend D., Bartlett, S., Cleaves, H., Demarest, H., Prabhu, A., Lunine, J. and Hazen R. (2023) On the roles of function and selection in evolving systems. *Proceedings of the National Academy of Sciences* 120, no. 43: e2310223120.
41) Smolin, L. (2013) *Time reborn: From the crisis in physics to the future of the universe*. HMH.

42) Kauffman, S. A. and Roli, A. (2021) The world is not a theorem. *Entropy* Vol 23, issue 11
*43)* Kauffman, S. A. and Roli, A. (2023) A third transition in science? Interface Focus13: 20220063. https://doi.org/10.1098/rsfs.2022.0063
44) Gould, S. J., and Vrba, E. S. (1982) Exaptation—a missing term in the science of form. *Paleobiology* 8, no. 1, 4-15.
45) Fraenkel, A., Bar-Hillel, Y. and Levy, A. (1973) *Foundations of set theory*. Elsevier.
46) Grattan-Guinness, I. (2000) *The search for mathematical roots, 1870-1940: logics, set theories and the foundations of mathematics from Cantor through Russell to Gödel*. Princeton University Press.
47) Skolem, Th. (1955) Peano's axioms and models of arithmetic. In *Studies in Logic and the Foundations of Mathematics*, vol. 16, pp. 1-14. Elsevier.
48) Cassidy, D. (1992) Heisenberg, uncertainty and the quantum revolution. *Scientific American* 266, no. 5, 106-113.
49) Exoplanet, Wikipedia
50) Krissansen-Totton, J., Thomson, M., Galloway, M., Fortney, J. (2022) Understanding planetary context to enable life detection on exoplanets and testing the Copernican Principle, *Nature Astronomy*, Vol 6, 189-198.
51) Stern, J., Makespin, C., Eigenbrode, J., and Mahaffy, P. Organic carbon concentrations in 3.5 billion year-old lacustrine mudstones of Mars, (2022), PNAS, 119 (27) e2201139119. https://doi.org/10.1073/pnas.2201139119
52) Xin, L. New Evidence Discovered that Saturn's Moon Could Support Life, (2023), Scientific American.
53) McKay, C., Smith, H. (2005), Possibilities for methanogeneic life in liquid methane on the surface of Titan. Icarus. 178 (1) 274. . Bibcode:2005Icar..178..274M. doi:10.1016/j.icarus.2005.05.018.
54) Wollrab, E., and Ott, A. (2018) A Miller–Urey broth mirrors the mass density distribution of all Beilstein indexed organic molecules. *New Journal of Physics* 20, no. 10 105003.119.
55) Lehman, N. E. and Kauffman, S. A, (2021), Constraint closure drove major transitions in the origins of life. *Entropy*, 23, 1, 105. https://doi.org/10.3390/e23010105
56) Carter C., Wills, P. (2017), Interdependence, Reflexivity, Fidelity, Impedence Matching, and the Evolution of Genetic Coding. *Molecular Biology and Evolution*, Vol 35, issue 2.
57) Kauffman, S. and Lehman, N. E. (2023) Mixed anhydrides at the intersection between peptide and RNA autocatalytic sets: evolution of biological coding, *Journal of the Royal Society Interface*. https://doi.org/10.1098/rsfs.2023.0009
58) Bordenave, G., Louis Pasteur (1822–1895). (2003) *Microbes and infection* 5, no. 6, 553-560.
59) Lazcano, A. (2016) Alexandr I. Oparin and the origin of life: a historical reassessment of the heterotrophic theory. *Journal of molecular evolution* 83, 214-222.
60) Tirard, S. (2017) JBS Haldane and the origin of life. *Journal of genetics* 96, no. 5, 735-739.
61) Miller, S. and Urey, H. (1959) Organic Compound Synthesis on the Primitive Earth. *Science* Vol 130 (3370) 245–51.

62) Macintyre, A. (2011) The impact of Gödel's incompleteness theorems on mathematics. *Kurt Gödel and the foundations of mathematics: Horizons of Truth* 3-25.